\documentclass{tran-l}
\theoremstyle{definition}

\theoremstyle{remark}

\numberwithin{equation}{section}
\begin{document}

\title[]
 {On The Geometrical and Physical Properties of Spherically Symmetric Non-Static Space-Times: Self-Similarity}

\author[ragab M. Gad]{ Ragab M. Gad}

\author[M. M. Hassan]{  M. M. Hassan}
\address{ Mathematics Department, Faculty of
 Science, Minia University, Minia, Egypt}

\email{}

\thanks{E-mail: ragab2gad@hotmail.com}

\thanks{}

%\subjclass{}

%\keywords{}

%\date{}

\dedicatory{}

%\commby{}
%%% ----------------------------------------------------------------------

\begin{abstract}
An exact solution of the Einstein field equations is found under the assumption of spherically
symmetry and the existence of one-parameter group of homothetic motions. This solution
has a singularity at $r = 0$, and has non-vanishing expansion, acceleration and shear. Tidal
forces in radial direction will not stretch an observer falling in this fluid. If the material is represented
by perfect fluid, we can verify that the solution coincides with stiff matter case. In this case, the solution
has zero expansion. Tidal forces in radial direction will not stretch an observer falling in this fluid and
they not squeeze him in transverse directions.\\


\end{abstract}

%%% ----------------------------------------------------------------------
\maketitle
%%% ----------------------------------------------------------------------
 \section{Introduction}

Many exact solutions of Einstein equations were found by
requiring certain symmetry. Many survey articles are given
to discuss the concept of these symmetries in general
relativity from the mathematical viewpoint, see for
example \cite{Hall96}, \cite{Hall98}.
In this paper we restricted our intention to study self-similarity,
which is one of the more important symmetry.

\par
A self-similar space-time is characterized by the existence of a homothetic Killing vector
field $H$ \cite{3}, \cite{5}. This vector satisfies the following equation
\begin{equation}\label{1.1}
\pounds_{H} g_{ab} = H_{a;b} + H_{b;a} = 2\Phi g_{ab},
\end{equation}
where $\pounds$ stands for the lie derivative, a semicolon for a covariant derivative and $\Phi$ is
constant. When $\Phi = 0$, $H$ is called a Killing vector field but when $\Phi$ is an arbitrary
function of coordinates, $H$ is called a Killing conformal vector field.

\par
The concept of self-similarity is largely documented
in the literature of general relativity, beginning with the
paper by Cahill and Taub \cite{3}, and followed by
important work by Eardley \cite{5}, \cite{E74}.
More recently Carter and Henriksen \cite{CH91} and
Sintes et al \cite{SBC01} have introduced the concept
of Kinematic self-similarity which is a generalization
of the homothetic case. Wainwright \cite{W00} has
compared the different mechanisms for asymptotic
self-similarity breaking in Bianchi universe.

\par
Several authors are studied the spherically symmetric solutions
admitting a self-similar motion orthogonal to the four-velocity
vector. Most of them have restricted their intention to the
solutions discovered by Gutman-Be'spalko \cite{6}, which are given
in another form by Wesson \cite{10}. These solutions are denoted
by GBW. Collins and Land \cite{4}, are studied these solutions as
well as a stiff equation of state. Sussman \cite{9} investigated
the properties
of them, and obtain interesting results. \\
By assuming an additional homothetic symmetry on
the spherically symmetric space-time
Barrete et al \cite{BOR98} have studied the
evolution of radiating and viscous fluid spheres.
They have hound a self-similar collapse which is
shear free and with a barotropic equation of state.

\par
The aim of this paper is to obtain an exact new form of non-static spherically symmetric solution,
which admits a self-similar motion orthogonal to the four velocity vector, and then study some of
its physical properties.\\
For the study of non-static spherically symmetric, we use the model given by \cite{11}
\begin{equation}\label{1.2}
ds^2 = \alpha (\nu , r)d\nu^2 + 2\beta (\nu , r)d\nu dr - r^2 (d\theta^2 + \sin^{2}\theta d\phi^2 ),
\end{equation}
where $\alpha$ and $\beta$ are positive functions of $\nu$ and $r$.
\par
 As pointed by Cahill and Taub \cite{3} and Bicknell and Henriksen \cite{1} (see also \cite{8}),
if the matter field is perfect fluid, then the only baratropic equation of state which is compatible
with self-similar is of the form
\begin{equation}\label{1.3}
p = k \rho,
\end{equation}
where $\rho$ is the total energy density, $p$ is the pressure and $k$ is a constant in the
rang $0 \leq  k \leq 1$. This equation of state is nevertheless physically consistent in the
whole rang of $k$. When $k = 0$, equation (\ref{1.3}) describes dust and $k = \frac{1}{3}$
gives the equation of state for radiation.
\par
In this paper we will show that the only case which is compatible with self-similar non-static
spherically symmetric space-time is $k = 1$, which is coincides with stiff-matter case. Moreover,
we show that this model contains shear, negative acceleration and zero expansion.

\par
The organization of the paper is as follows: in section 2, we assume that the model (\ref{1.2})
admits homothetic symmetry. Under the assumption of self-similarity, we derive an exact solution.

In section 3, we assume, in addition to self similarity, that the material is represented by perfect fluid.
We will show that the only baratropic equation of state, which is compatible with self-similar non-static
spherically symmetric space-time is $p = \rho$.\\
Finally in section 4, the physical structures of the solutions mentioned in section 2 are discussed.
We have discussed the  expansion,
acceleration and shear for these solutions.
The singularities and tidal forces are also investigated.

\section{Homothetic Equations}
In this section, we obtain the homothetic symmetry for the model (\ref{1.2}). We drive new model of
non-static spherically symmetric admits homothetic symmetry. The homothetic Killing equations (\ref{1.1})
for the model (\ref{1.2}) are reduced to the following system of equations
\begin{equation} \label{2.1}
H^{0}_{r} = 0,
\end{equation}
\begin{equation} \label{2.2}
\beta H^{0}_{\theta} - r^{2}H^{2}_{r} = 0,
\end{equation}
\begin{equation} \label{2.3}
\beta H^{0}_{\phi} - r^{2}\sin^{2}\theta H^3_{r} = 0,
\end{equation}
\begin{equation} \label{2.4}
\beta H^{0}_{\nu} + \beta_{\nu} H^{0} + \beta H^{1}_{r} + \beta_{r} H^{1} = 2\Phi\beta,
\end{equation}
\begin{equation} \label{2.5}
H^{2}_{\theta} + \frac{1}{r} H^{1} = \Phi,
\end{equation}
\begin{equation} \label{2.6}
H^{2}_{\phi} + \sin^{2}\theta H^{3}_{\theta} = 0,
\end{equation}
\begin{equation} \label{2.7}
\beta H^{1}_{\theta} - r^{2} H^{2}_{\nu} + \alpha H^{0}_{\theta} = 0,
\end{equation}

\begin{equation} \label{2.8}
H^{3}_{\phi} + \cot\theta H^{2} + \frac{1}{r} H^{1} = \Phi,
\end{equation}
\begin{equation} \label{2.9}
\beta H^{1}_{\phi} - r^{2}\sin^{2}\theta H^{3}_{\nu} + \alpha H^{0}_{\phi} = 0,
\end{equation}
\begin{equation} \label{2.10}
\beta H^{1}_{\nu} + \frac{\alpha_{r}}{2} H^{1} + \frac{\alpha_{\nu}}{2} H^{0} + \alpha H^{0}_{\nu} = \Phi\alpha,
\end{equation}
where subscripts denote partial differentiation. The above equations are a coupled system of first order
linear partial differential equations for the homothetic vector ${\bf{H}} = ( H^0, H^1, H^2, H^3 )$.
\par
Unfortunately, it's not easy to find the homothetic symmetry vector without making any assumption
about the form of homothetic vector.\\
We shall further restrict the vector field ${\bf{H}}$, by demanding
\begin{equation} \label{2.11}
H^{a}u_{a} = 0,
\end{equation}
where $u^{a} = \delta^{a}_{0}\alpha^{-\frac{1}{2}}$ is the four-velocity of the fluid, $u^{a}u_{a} = 1$. \\
Then, as a consequence of the spherical symmetry (it is possible to put $\theta = \frac{\pi}{2} = $ constant)
and from equation (\ref{2.11}), we have
$$
H^0 = H^2 = H^3 = 0.
$$
The system of equations (\ref{2.1})-(\ref{2.10}) are reduced to the following equations
\begin{equation} \label{2.12}
\beta H^{1}_{r} + \beta_{r} H^{1} = 2\Phi\beta,
\end{equation}
\begin{equation} \label{2.13}
H^{1} = \Phi r,
\end{equation}
\begin{equation} \label{2.14}
\beta H^{1}_{\nu} + \frac{\alpha_{r}}{2} H^{1} = \Phi \alpha.
\end{equation}
Using equation (\ref{2.13}) in equations (\ref{2.12}) and
(\ref{2.14}), we get the unknown functions $\alpha$ and $\beta$ in
the following form
\begin{equation} \label{2.15}
\alpha = r^{2} h(\nu), \qquad \beta = rf(\nu),
\end{equation}
where $h(\nu)$ and $f(\nu)$ are arbitrary functions.\\
Now, we can write down the line element (\ref{1.1}) in terms of the functions $h(\nu)$ and $f(\nu)$ as follows
\begin{equation} \label{2.16}
ds^2 = r^{2}h(\nu)d\nu^2 + 2r f(\nu) d\nu dr - r^{2} (d\theta^2 + \sin^{2}\theta d\phi^2 ).
\end{equation}

\section{Einstein Field equations}
In addition to self-similarity, we assume that the matter is represented by perfect fluid,
i.e., the Einstein field equations $G_{ab} = - \kappa T_{ab}$, are satisfied with the energy
momentum tensor
$$
T_{ab} = (\rho + p)u_{a}u^{a} - pg_{ab}.
$$
For the line element (\ref{2.12}) the Einstein field equations can be written as follows
\begin{equation} \label{3.1}
\frac{2}{r^2} = \kappa (\rho + p)\frac{f^{2}(\nu)}{h(\nu)},
\end{equation}
\begin{equation} \label{3.2}
\frac{1}{r^2} - \frac{h(\nu)}{r^{2}f^{2}(\nu)} = \kappa\rho,
\end{equation}
\begin{equation} \label{3.3}
\frac{h(\nu)}{r^{2}f^{2}(\nu)} = \kappa p,
\end{equation}
\begin{equation} \label{3.4}
\frac{1}{r^2} - \frac{2\dot{f}(\nu)}{r^2 h(\nu)} + \frac{\dot{h}(\nu)}{r^{2}f^{2}(\nu)} = \kappa\rho,
\end{equation}
where dot denotes differentiation with respect to $\nu$.\\
Using equations (\ref{3.2}) and (\ref{3.3}) in equation (\ref{3.1}), we get the relation
between the two functions $f(\nu)$ and $h(\nu)$ in the form
\begin{equation} \label{3.5}
f^{2}(\nu) = 2h(\nu).
\end{equation}
According to equation (\ref{3.5}), we see that equation (\ref{3.4}) is equivalent to equation (\ref{3.2}).\\
Now, from equations (\ref{3.1}) - (\ref{3.5}), we can obtain the expressions for the pressure and density in the form
\begin{equation} \label{3.6}
p = \rho = \frac{1}{2\kappa r^2}.
\end{equation}
This equation can be viewed as an equation of state of stiff matter.\\
In the case of perfect fluid, the line element (\ref{2.16}) can be expressed in the form
\begin{equation} \label{3.7}
ds^2 = rf(\nu)\big( \frac{1}{2}rf(\nu)d\nu^2 + 2d\nu dr \big) - r^2 (d\theta^2 + \sin^{2}\theta d\phi^2).
\end{equation}

\section{Physical Properties}
In this section we discuss some  physical structures of our obtained  solutions (\ref{2.16}) and (\ref{3.7}).
%\numberwithin{equation}{subsection}
\subsection{Singularities}
The Kretschmann scalar ($K \equiv R_{abcd}R^{abcd}, \ R_{abcd}$
the Riemann tensor) for the metric (\ref{2.16}) reduces to
$$
K \equiv \frac{1}{r^4 f^{4}(\nu)}\big( 3h^{2}(\nu) - 4h(\nu) \frac{df(\nu)}{d\nu} + 2f(\nu)\frac{dh(\nu)}{d\nu} \big),
$$
and so the metric is scalar-polynomial (SP) singular along $r = 0$, because $K$ diverges at this point. \\
For the perfect fluid solution (\ref{3.7}), using equation (\ref{3.5}), the Kertschmann scalar reduces to
$$
K = \frac{3}{4r^4},
$$
so the metric is also scalar-polynomial singular along $r = 0$. From another point of view the point
$r = 0$ is a singular for the solution (\ref{3.7}) where the density necessarily diverges. It is clear from
equation (\ref{3.6}) that the value of the density is infinite at this point.

\subsection{Kinematic of the Velocity Field}

We now drive the kinematic quantities of the solution (\ref{2.16})\\
The expansion scalar $\Theta = u^{a}_{;a}$, which determines the volume behavior of the fluid, is given by
$$
\Theta = \frac{1}{r\sqrt{h(\nu}}\big(\frac{\dot{f}(\nu)}{f(\nu)} - \frac{\dot{h}(\nu)}{2h(\nu)}\big).
$$
For the acceleration we find
$$
\dot{u}_{a} u_{a;b}u^{b} = \big( \frac{f(\nu)}{\sqrt{h(\nu)}}\Theta - \frac{1}{r}\big) \delta^{1}_{a}.
$$
The shear tensor, $\sigma_{ab} = u_{(a;b)} + \dot{u}_{(a}u_{b)} - \frac{1}{3}\Theta h_{ab}$,
determines the distortion arising in the fluid flow leaving the volume invariant.
The non-zero components of the shear tensor are

$$
\sigma_{00} = - \frac{1}{3}r^{2}h(\nu)\Theta, \qquad \sigma_{10} = \frac{1}{3}rf(\nu)\Theta,
$$
$$
\sigma_{11} = \frac{f^{2}(\nu)}{h(\nu)}\Theta - \frac{2f(\nu)}{r\sqrt{h(\nu)}},
$$
$$
\sigma_{22} = \frac{1}{3}r^{2}\Theta, \qquad \sigma_{33}=  \frac{1}{3}r^{2}\Theta\sin^{2}\theta,
$$
and the shear scalar, $\sigma^{2} = \frac{1}{2}\sigma_{ab}\sigma^{ab}$, is given by
$$
\sigma^2 = \frac{19}{18}\Theta^2 - \frac{8\sqrt{h(\nu)}}{3rf(\nu)}\Theta + \frac{2h(\nu)}{r^{2}f^{2}(\nu)}.
$$
\par
In the case of perfect fluid solution (solution (\ref{3.7})), using (\ref{3.5}), the expansion scalar is zero,
the acceleration is given by \, $\dot{u}_{a} = - \frac{1}{r}\delta^{1}_{a}$. For the shear tensor, the
non-vanishing component is \, $\sigma_{11} = - \frac{2\sqrt{2}}{r}$, \, and the shear scalar is given
by \, $\sigma^{2} = \frac{1}{r^2}$.
\subsection{Tidal Forces}
The components of the Riemann curvature tensor $R^{a}_{bcd}$, which describe tidal forces
(relative acceleration) between two particles in free fall, are the components
\, $R^{i}_{0j0}$, ($i, j = 1, 2, 3$), \cite{7}.\\
For the line element (\ref{2.16}), we obtain
$$
R^{1}_{010} = 0,
$$
and the only non-vanishing relevant components are
$$
R^{2}_{020} = R^{3}_{030} = \frac{h^{2}(\nu)}{f^{2}(\nu)} + \frac{\dot{h}(\nu)}{2f(\nu)}
- \frac{h(\nu)\dot{f}(\nu)}{f^{2}(\nu)}.
$$

Then the equations of geodesic deviation (Jacobi equations), which connected between
the behavior of nearby particles and curvature, are reduce to the following equations
\begin{equation}\label{4.1}
\frac{D^{2}\zeta^{r}}{d\tau^2} = 0,
\end{equation}
\begin{equation}\label{4.2}
\frac{D^{2}\zeta^{\theta}}{d\tau^2} =- \frac{1}{r^{2}h(\nu)}\big( \frac{h^{2}(\nu)}{f^{2}(\nu)} + \frac{\dot{h}(\nu)}{2f(\nu)}
- \frac{h(\nu)\dot{f}(\nu)}{f^{2}(\nu)}\big) \zeta^{\theta} ,
\end{equation}
\begin{equation}\label{4.3}
\frac{D^{2}\zeta^{\phi}}{d\tau^2} =- \frac{1}{r^{2}h(\nu)}\big( \frac{h^{2}(\nu)}{f^{2}(\nu)} + \frac{\dot{h}(\nu)}{2f(\nu)}
- \frac{h(\nu)\dot{f}(\nu)}{f^{2}(\nu)}\big) \zeta^{\phi} ,
\end{equation}
where $\zeta^r , \ \zeta^{\theta} , \ \zeta^{\phi}$ are the components of Jacobi vector field.\\
Hence, equation (\ref{4.1}) indicates tidal forces in radial direction will not stretch an observer
falling in this fluid. If $2h^{2}(\nu) + f(\nu)\dot{h}(\nu) - 2h(\nu)\dot{f}(\nu)$ is positive (negative),
then the equations (\ref{4.2})and (\ref{4.3}) are indicate a pressure or compression
(tension or stretching) in the transverse directions.
\par
In the case of self-similar perfect fluid solution (\ref{3.7}), after using (\ref{3.5}) in equations
(\ref{4.1}) - (\ref{4.3}), the tidal forces in radial direction will not also stretch the observer,
and they will not squeeze him in the transverse directions.

\section*{Conclusions}

We have assumed an additional symmetry to the spherically symmetric space-time,
homothetic motion, to obtain non-static solution. This solution has non-vanishing
expansion, acceleration and shear. Another solution has been derived by assuming
the matter in this fluid is represented by perfect fluid. In this case the solution has zero
expansion. For more physically properties of the obtained solutions, we have discussed
their singularities and the behavior of tidal forces in both of them.

\end{document}